\documentclass{article}
\usepackage{arxiv}

\usepackage[utf8]{inputenc}
\usepackage[T1]{fontenc}

\usepackage{hyperref}
\usepackage{url}
\usepackage{booktabs}
\usepackage{amsfonts,amsmath,amssymb}
\usepackage{nicefrac}
\usepackage{microtype}
\usepackage{graphicx}
\usepackage{xcolor}
\usepackage{caption}
\usepackage{subcaption}
\usepackage{tabularx}
\usepackage{makecell}
\usepackage{pgfplots}
\usepackage{pgfplotstable}
\usepackage{fvextra}

\pgfplotsset{compat=1.18}

\hypersetup{
    colorlinks=true,
    linkcolor=blue,
    citecolor=blue,
    urlcolor=blue
}

\expandafter\def\expandafter\UrlBreaks\expandafter{%
  \UrlBreaks\do\/\do\*\do\-\do\~\do\'\do\"\do\-%
}

\title{CLiVR: Conversational Learning System in Virtual Reality with AI-Powered Patients}

\author{
Akilan Amithasagaran \\
The University of Texas at Tyler\\
Tyler, TX, USA\\
\texttt{aamithasagaran@patriots.uttyler.edu} \\
\And
Sagnik Dakshit \\
The University of Texas at Tyler\\
Tyler, TX, USA\\
\texttt{sdakshit@uttyler.edu} \\
\And
Bhavani Suryadevara \\
The University of Texas at Tyler\\
Tyler, TX, USA\\
\texttt{Bhavani.Suryadevara@uttyler.edu} \\
\And
Lindsey Stockton \\
The University of Texas at Tyler\\
Tyler, TX, USA\\
\texttt{lindsey.stockton@uttyler.edu}
}

\begin{document}
\maketitle

\begin{abstract}
Simulations constitute a fundamental component of medical and nursing education and traditionally employ standardized patients (SP) and high-fidelity manikins to develop clinical reasoning and communication skills. However, these methods require substantial resources, limiting accessibility and scalability. In this study, we introduce \textbf{CLiVR}, a Conversational Learning system in Virtual Reality that integrates large language models (LLMs), speech processing, and 3D avatars to simulate realistic doctor–patient interactions. Developed in Unity and deployed on the Meta Quest 3 platform, CLiVR enables trainees to engage in natural dialogue with virtual patients. Each simulation is dynamically generated from a syndrome-symptom database and enhanced with sentiment analysis to provide feedback on communication tone. Through an expert user study involving medical school faculty (n=13), we assessed usability, realism, and perceived educational impact. Results demonstrated strong user acceptance, high confidence in educational potential, and valuable feedback for improvement. CLiVR offers a scalable, immersive supplement to SP-based training.
\end{abstract}

\section{Introduction}
Effective communication between physicians and patients is fundamental to clinical competence, and is a primary focus of medical and nursing education. To facilitate the development of these skills among learners, educational institutions frequently utilize simulation-based training methods, such as standardized patients (SPs) and high-fidelity manikins. SPs are trained actors depicting clinical scenarios. This is particularly beneficial for enhancing interpersonal skills, empathy, and diagnostic dialogue for tasks such as patient triage. Conversely, manikins are programmable physical simulators that replicate physiological responses, including heart and lung sounds, palpable pulses, and verbal cues, all of which are controlled through software interfaces. Although both methods have shown similar effectiveness in terms of knowledge acquisition, SP-based training is often more effective in improving communication skills. Furthermore, highly realistic simulation environments better equip students for clinical practice by offering safe and repeatable opportunities to navigate complex and diverse scenarios.

These traditional modalities in medical education exhibit inherent limitations of being costly, resource intensive, and challenging to scale, particularly in environments with restricted access to trained actors or advanced simulation laboratories. In response, emerging technologies such as virtual reality (VR) and artificial intelligence (AI), particularly large language models (LLMs), present promising alternatives \cite{ali2025role}. VR offers immersive and engaging environments that replicate clinical settings, enabling students to practice without physical constraints. Concurrently, the recent proliferation of LLMs, including ChatGPT, LLaMA, and Gemma, has demonstrated AI's capacity to generate coherent, context-aware, and responsive dialogue \cite{yankouskaya2025can}. These models can now emulate specific personas \cite{salminen2024deus}, adhere to structured clinical prompts, and adapt dynamically to user input capabilities that are particularly well suited for simulating patient interactions. The convergence of VR and LLMs represents a significant opportunity to reconceptualize the teaching and practice of medical communication. Unlike static or scripted simulations, AI-driven virtual patients can simulate a wide array of emotional states, communication barriers, and clinical scenarios, thereby providing learners with more diverse and nuanced practice opportunities. This includes the ability to represent a broad spectrum of medical conditions ranging from common ailments to complex, multisystem disorders, while also incorporating psychosocial contexts such as mental health, substance use, and chronic illness management. Moreover, simulations can be designed to expose learners to sensitive and high-stakes conversations involving race, gender identity, sexual orientation, socioeconomic status, disability, and other factors that affect care equity. However, these interpersonal competencies remain among the most challenging to assess and enhance through lectures or written assessments. These features are particularly impactful in fostering culturally competent communication and reducing implicit bias in clinical encounters. By modeling interactions with patients from underrepresented or vulnerable populations, the system supports experiential learning in areas that are often difficult to teach using conventional methods. This is particularly critical in communication training, where empathy, tone, timing, and adaptability are essential for building trust, gathering accurate histories, and conveying difficult information with sensitivity \cite{elendu2024impact}. 

In this study, we present \textbf{CLiVR}, a \textit{Conversational Learning system in Virtual Reality} engineered to replicate authentic doctor–patient interactions through AI-driven 3D avatars for educational purposes. The system utilizes LLMs to play the role of a patient and is capable of generating a variety of symptom-specific medical scenarios. These scenarios are derived from a meticulously curated syndrome-symptom knowledge base and are delivered through speech-based interactions facilitated by real-time transcription, text-to-speech synthesis, and realistic lip-sync animations of virtual patients. Furthermore, the system analyzes sentiment of user response to offer feedback on emotional tone and communication style. The development of CLiVR necessitated addressing several technical challenges: (a) designing VR application with expressive 3D patient avatars, (b) grounding LLM responses for precise and consistent clinical role-play using retrieval-augmented generation (RAG), (c) synchronizing avatar lip movements with synthesized speech using 3D blendshapes, and (d) analyzing trainee sentiments in medical context where neutral communication is standard during interactions to enhance reflective learning. To assess the feasibility and instructional value of our system, we conducted a exploratory user study involving medical educators. The findings indicate high user satisfaction and underscore the potential of CLiVR as a scalable, engaging, and pedagogically significant tool for communication training in professional health education.

Our key contributions can be summarized as:
\begin{itemize}
    \item This study introduces CLiVR, a \textbf{ new modular VR-based simulation platform for clinical communication training.} This platform integrates LLMs, speech processing, sentiment analysis, and 3D avatar interaction to support immersive doctor–patient simulations with near real-time performance of 1.4 seconds latency (Meta Quest 3).

    \item \textbf{Support for emotionally and responsive training scenarios:} CLiVR enables trainees to practice communication and empathy in diverse simulated clinical and sociocultural variations through sentiment analysis.

    \item \textbf{Empirical validation with medical trainees and faculty:} Through a mixed-methods exploratory study with medical experts (\(n=13\)), we demonstrate the feasibility, usability, and educational relevance of CLiVR reporting high engagement and learning potential.
\end{itemize}

\section{Related Works}
\label{sec:related}
The integration of conversational learning systems with AI-powered VR environments is transforming educational methodologies across diverse domains, including healthcare and general education. These approaches leverage AI to create personalized, interactive, and immersive learning experiences that go beyond traditional pedagogical models to deliver personalized learning. Intelligent tutoring and adaptive learning management systems exemplify this shift from one-size-fits-all to individualized education, improving engagement and outcomes \cite{alashwal2024empowering}.

Several studies have explored the use of AI-powered virtual tutors equipped with natural language processing (NLP) capabilities to enable real-time interactive dialogues as AI tutors \cite{rathika2024developing}, particularly in medical training such as massage therapy training \cite{wang2025theraquest}, surgical simulations \cite{javaheri2025llms}, nursing education \cite{hu2025nurse}, and communication training for medical responders \cite{gutierrez2024integrating}. However, the nuances of patient triaging in VR grounded in medical knowledge has not been explored to the best of our knowledge.

In parallel to advances in AI, VR has emerged as a powerful tool for immersive healthcare education, VR enables repeated, risk-free practice of clinical procedures in simulated settings, which is critical for developing high-stake decision-making skills \cite{guraya2024transforming}. The integration of AI-powered simulations and VR also provides a safe and controlled environment. Kenny et al. \cite{kenny2024virtual} discussed the applications of AI powered SPs, which we leverage in the development of our proposed CLiVR system. Mergen et al., \cite{mergen2023immersive} proposed a platform for face-to-face interactions with virtual patients, simulating clinical scenarios that include highly realistic pathologies in customizable situational contexts. However, their implementation details have not been published for reproducibility. Other studies have evaluated LLMs for clinical dialogue generation \cite{kapadia2024evaluation} and domain-specific clinical reasoning \cite{borg2024enhancing}; however, integration with immersive VR remains limited. Maslych et al. \cite{maslych2025avatars} conducted a VR pilot study applying LLMs to multiple conversational avatars, demonstrating the feasibility of real-time dialogue and evaluating user experience with interface features such as status indicators. Their study offers insights into avatar interaction design within VR-based conversational systems but does not focus on design factors or grounding LLM driven conversations to medical context. Additionally their work does not take into account sentiment for empathic learning which is critical for medical training systems.

CLiVR advances prior work by combining capabilities that have not been previously combined in medical communication training platforms. Comparison with commercial VR tools such as Oxford Medical Simulation, PCS Spark, and Body Interact are illustrated in Section \ref{sec:Comparison}. CLiVR supports open-ended, symptom-constrained interactions grounded in a curated syndrome–symptom database, while preserving clinical realism against commercial pre-scripted systems. Leveraging LLMs with structured prompt engineering to create context-aware, persona-driven, and emotionally responsive virtual patients, it delivers real-time empathy feedback through integrated sentiment analysis, enabling immediate reflection and adaptation.

\section{CLiVR System Overview}
\label{sec:CliVR}

\begin{figure*}[!t]
    \centering
    \includegraphics[width=\textwidth]{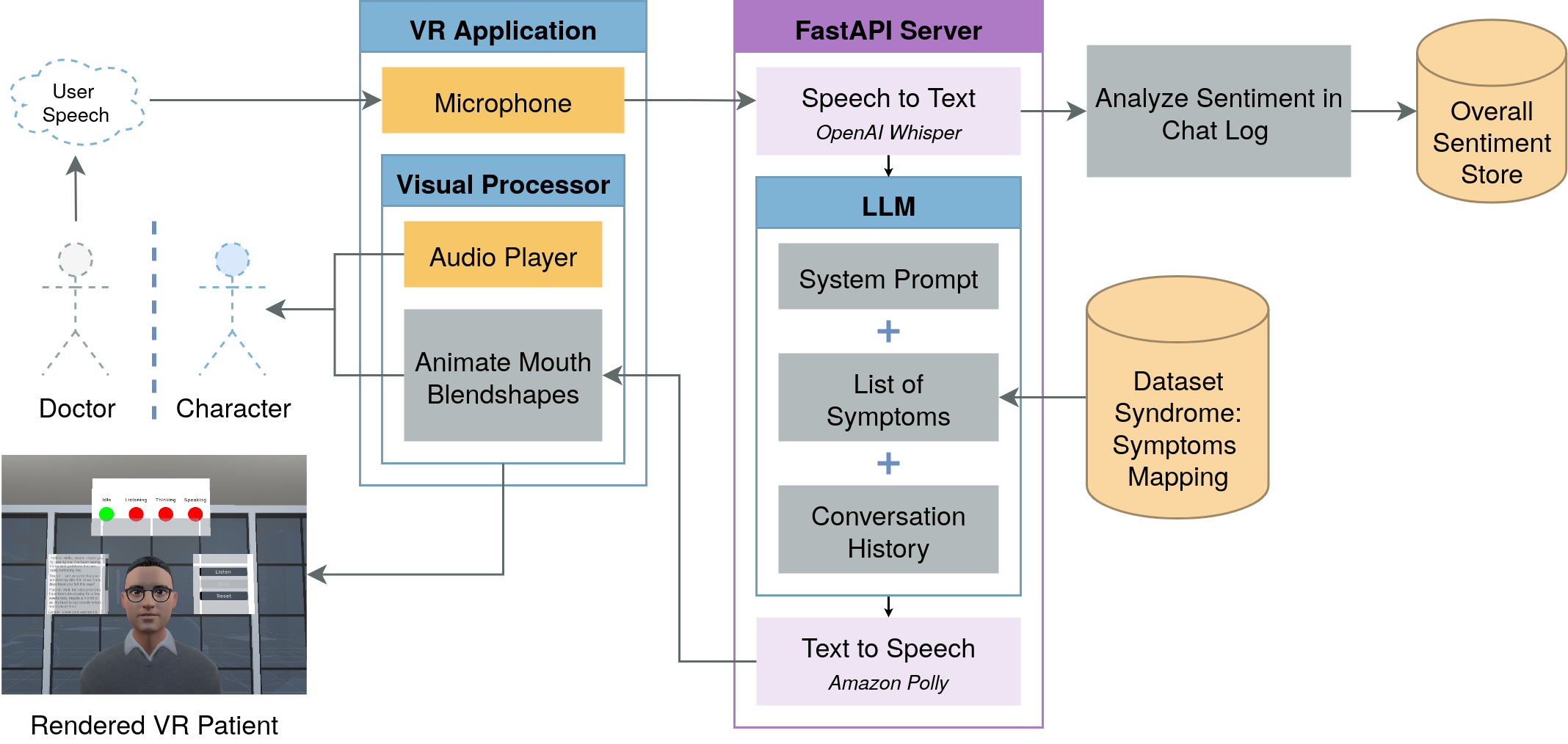}
    \caption{End-to-end system architecture of the CLiVR application}
    \label{fig:clivr_architecture}
\end{figure*}

In this section, we discuss the development and deployment of our CLiVR application with a focus on the system architectural components and data set to ground medical knowledge of LLMs.

\subsection{Syndrome and Symptom Data Curation}
\label{sec:dataset}
To ensure medical accuracy and minimize hallucinations, CLiVR grounds each patient scenario in a curated syndrome–symptom knowledge base that constrains the LLM to plausible clinical behavior merging two open-source datasets:

\begin{itemize}
    \item \textbf{Mendeley Disease Dataset}~\cite{nayaki2024clinical}: 4,961 disease–symptom associations compiled from public medical resources.
    \item \textbf{Columbia University Disease–Symptom Knowledge Base}~\cite{wang2008automated}: 134 associations extracted from discharge summaries at New York Presbyterian Hospital.
\end{itemize}

The merged dataset provides 5{,}095 syndrome–symptom pairs. At the start of each simulation, the system randomly selects one syndrome and its symptoms and injects them into the LLM prompt, guiding it to role-play a patient with those symptoms. This ensures experience are distinct, diverse and medically consistent.

\subsection{System Architectural Components}

The CLiVR system adopts a modular client–server architecture integrating a Unity-based VR front end with a lightweight FastAPI backend for language, speech, and sentiment processing (Figure~\ref{fig:clivr_architecture}). This architecture supports dynamic speech-driven scenarios in virtual environments that emulate real-world clinical consultations, providing a scalable and pedagogically relevant alternative to traditional standardized patient- or mannequin-based simulations. This client–server configuration facilitates deployment on standalone headsets, such as Meta Quest 3, while delegating language generation and audio synthesis tasks to a lightweight, extensible server.

\noindent \textbf{Front-End Module:}
The VR client captures trainee speech through the headset microphone, plays patient responses, and renders synchronized 3D facial animations within Unity’s real-time visual pipeline. Virtual patients are created using \textit{Ready Player Me}, which generates detailed 3D avatars imported directly into the Unity environment. To align facial movements with spoken audio, CLiVR uses \textit{uLipSync} an open-source library that calculates Mel-Frequency Cepstrum Coefficients (MFCC) to detect phonemes in the audio stream. This phoneme information is then used to adjust the avatar's mesh blend shapes dynamically, ensuring accurate synchronization of audio and visual cues. CLiVR incorporates a four-state status indicator  (Idle, Listening, Thinking, Speaking) inspired by Maslych et al. \cite{maslych2025avatars}, providing visual cues that reinforce user trust in system responsiveness.

\noindent \textbf{Back-End Module:}
The processing server performs the following sequence:
\begin{enumerate}
    \item \textit{Speech-to-Text Conversion:} Implemented using the OpenAI Whisper model for accurate, low-latency transcription.
    \item \textit{LLM Response Generation:} The transcribed input is passed to a large language model (e.g., Gemini 2.0-Flash) along with (i) a structured system prompt, (ii) the selected syndrome–symptom list, and (iii) short-term conversation memory. These elements constrain dialogue to clinically consistent and contextually coherent patient behavior.
    \item \textit{Text-to-Speech Synthesis:} Generated text is vocalized using Amazon Polly’s neural voices and returned to the VR client.
\end{enumerate}
The complete exchange is logged for sentiment analysis and post-session reflection. This modular workflow enables extensibility across different models or cloud endpoints without altering the VR application.

\noindent \textbf{Sentiment Analysis:}
 Sentiment analysis module enables evaluation of each doctor statement’s emotional tone throughout the interaction, supporting educational goals such as assessing empathy and communication style. Each transcribed message is classified as \textit{positive}, \textit{negative}, or \textit{neutral} using natural language models. The models are selected through unsupervised fine-tuning and testing on synthetic datasets for medical context-specific emotion detection, as illustrated in Section \ref{sec:sentiment}. This provides insights into trainees’ emotional responses to clinical scenarios, guiding improvements in communication skills, and adaptive feedback mechanisms. 

\noindent \textbf{System Latency:}
To ensure conversational realism, latency was profiled across four modules: speech recognition, LLM generation, speech synthesis, and sentiment inference. The mean round-trip delay per turn was approximately \textbf{1.35 s}, dominated by LLM generation (0.56 s) and followed by text-to-speech (0.24 s) and sentiment analysis (0.30 s). Whisper transcription remained minimal (0.14 s). These results confirm that CLiVR achieves near real-time responsiveness suitable for immersive medical training.

\subsection{LLM Roleplay and Prompt Engineering}
\label{sec:prompt}

LLMs enable coherent, context-aware dialogue generation and can adopt specific personas without task-specific fine-tuning. CLiVR leverages these zero-shot roleplay capabilities through structured prompt engineering~\cite{dietrich2025prompt} to simulate realistic medical interactions. Each session initializes the model with a structured system prompt that instructs it to assume the role of a patient constrained by the selected symptoms and previous conversational turns. This design allows trainees to experience diverse patient personalities and clinical presentations, thereby supporting both diagnostic reasoning and triage practice.

At the start of each session, the LLM is initialized with a concise system prompt that:
\begin{enumerate}
    \item Defines the \textit{patient role} (demographics, personality, communication style).
    \item Specifies \textit{symptom constraints} and medical context.
    \item Instructs the model to maintain \textit{temporal and logical consistency} across turns by referencing prior conversational history.
    \item Enforces \textit{empathy and cultural appropriateness}, while allowing variability in affective tone to match the case design.
\end{enumerate}

This approach enables trainees to encounter a broad spectrum of patient personas and clinical presentations, supporting their skills in \textit{diagnostic reasoning}, \textit{ history taking }, and \textit{triage decision-making}. It also facilitates consistent scenario reproduction while allowing for subtle behavioral variations between runs, mimicking the unpredictability of real patients. To achieve this, a representative system prompt encodes patient traits, symptom boundaries, and interaction rules that guide the LLM to remain in character, speak naturally, and express appropriate affect. These structured textual constraints preserve both clinical accuracy and emotional realism, ensuring that each simulation feels authentic while remaining reproducible across sessions.

\section{Sentiment Analysis of Medical Conversation}
\label{sec:sentiment}

Within the CLiVR architecture, a dedicated sentiment module estimates the emotional tone of the trainee responses during simulations. This section details the workflow for developing that module and justifies the choice of models based on their accuracy and effectiveness in classifying sentiments in medical conversations.

\subsection{Datasets}
Three complementary sources were used: (i) GoEmotions \cite{demszky2020goemotions}, a public corpus of 58{,}009 Reddit comments; (ii) a balanced, domain-specific corpus of 1{,}500 synthetic doctor utterances (500 per class) created with Gemini-2.0-Flash; and (iii) anonymized user-study chat logs comprising 427 clinical conversations collected from 15 licensed medical professionals interacting with virtual patients via the CLiVR system. Each dataset served a distinct role in the workflow. The GoEmotions corpus was used to fine-tune the BERT-family models for supervised sentiment classification. The synthetic doctor-utterance dataset, served as a controlled benchmark to evaluate all models under identical conditions. Finally, the unlabeled clinician chat logs supported deployment validation, where entropy and inter-model agreement quantified consistency in real-world conditions lacking reference labels.

\begin{table}[!htbp]
\centering
\begin{tabularx}{\textwidth}{|X|cccc|cccc|}
\hline
\textbf{Model} &
\multicolumn{4}{c|}{\textbf{GoEmotions}} &
\multicolumn{4}{c|}{\textbf{LLM-Generated Synthetic Data}} \\
\cline{2-9}
 & \textbf{Acc} & \textbf{Prec} & \textbf{Rec} & \textbf{F1} & \textbf{Acc} & \textbf{Prec} & \textbf{Rec} & \textbf{F1} \\
\hline
roberta-base            & 0.746 & 0.744 & 0.746 & 0.744 & 0.713 & 0.788 & 0.713 & 0.696 \\
bert-base-uncased       & 0.748 & 0.746 & 0.748 & 0.747 & 0.691 & 0.773 & 0.691 & 0.678 \\
distilbert-base-uncased & 0.739 & 0.738 & 0.739 & 0.738 & 0.699 & 0.765 & 0.699 & 0.688 \\
gpt-4o-mini             & --    & --    & --    & --    & 0.853 & 0.875 & 0.853 & 0.849 \\
gemma3                  & --    & --    & --    & --    & 0.798 & 0.850 & 0.798 & 0.789 \\
gemma3n                 & --    & --    & --    & --    & 0.853 & 0.875 & 0.853 & 0.846 \\
mistral                 & --    & --    & --    & --    & 0.715 & 0.774 & 0.715 & 0.699 \\
\hline
\end{tabularx}
\caption{Performance on GoEmotions and LLM-generated doctor-conversation benchmark}
\label{tab:sentiment-results}
\end{table}

\subsection{Sentiment Evaluation}
Seven classifiers were evaluated: three fine-tuned BERT variants (\texttt{bert-base-uncased}, \texttt{roberta-base}, \texttt{distilbert-base-uncased}) and four lightweight large language models (\texttt{gpt-4o-mini}, \texttt{gemma3}, \texttt{gemma3n}, \texttt{mistral}). 
BERT models were fine-tuned on GoEmotions using standard transformer settings (Adam optimizer, learning rate $2{\times}10^{-5}$, two epochs, batch size 16), and the best validation checkpoint was retained. 
Each model was then evaluated on the 1{,}500-sentence doctor benchmark using standard sentiment classification metrics namely accuracy, precision, recall, and F1 score. 
Finally, all models were applied to unlabeled clinician logs to examine inter-model agreement (Cohen’s $\kappa$) and class-distribution entropy, indicating whether neutrality bias or mode collapse occurred. 
The LLMs were instruction prompted to classify each doctor's response as \textit{positive}, \textit{neutral}, or \textit{negative}, focusing strictly on tone, empathy, and reassurance.

\subsection{Results and Discussion}
The empirical evaluation yielded three principal findings. First, within the fine-tuned family, \texttt{bert-base-uncased} achieved the strongest aggregate performance on the GoEmotions test split, posting accuracy 0.748 and macro-F1 0.747 (Table \ref{tab:sentiment-results}). Second, among the lightweight large-language models, \texttt{gpt-4o-mini} and \texttt{gemma3n} attained the highest overall accuracy at 0.853 and led every per-class metric except negative-class recall, where \texttt{gemma3} edged ahead. These models boosted positive-class recall to at least 0.63, compared to 0.43-0.46 for the BERT variants, confirming their superior sensitivity to supportive or empathic phrasing. \texttt{mistral}, although more efficient than larger LLMs, was not chosen due to its comparatively weaker performance across sentiment classes. 

The first two datasets supported quantitative evaluation using labeled benchmarks, whereas the third unlabeled clinician chat logs was used solely for validation. Because these logs lacked sentiment ground truth, standard performance metrics such as accuracy or F1-score could not be computed. Instead, inter-model agreement (Cohen’s $\kappa$) and class-distribution entropy were analyzed to assess consistency and neutrality bias in realistic doctor–patient interactions. Low entropy values (0.38–0.83 bits) across all models indicated a predominance of neutral predictions, aligning with the expected tone of professional medical communication. Accordingly, these measures were excluded from the benchmark tables but served as complementary validation in real-world conversational data.

Taken together, these observations guided the final deployment decision. Although a hybrid configuration remains feasible, \texttt{gemma3n} was selected as the production model for its balance of performance and efficiency on the CLiVR workstation hardware. \texttt{GPT-4o-mini} remains accessible as an optional cloud resource for periodic, high-resolution transcript reviews, while all real-time feedback in the current system is generated locally by \texttt{gemma3n}.

\section{CLiVR Clinical Evaluation}
\label{sec:userstudy}

This section presents our mixed-methods expert feedback framework alongside both qualitative and quantitative evaluations of CLiVR. Additionally, we address the current limitations and propose future enhancements for AI-VR systems in medical education.

\subsection{Study Design}

To evaluate the clinical relevance, educational value, and usability of the CLiVR system, a mixed-methods study was conducted with medical school faculty. Participants utilized the CLiVR virtual simulation system via Meta Quest 3 headsets to engage in LLM-driven doctor–patient interactions, followed by a comprehensive questionnaire. Each participant engaged in a pre-session debriefing followed by one or more simulated clinical case studies with virtual patients powered by a LLM, and a post-session questionnaire. The scope was to assess not only the system's technical capabilities but also its educational impact on communication, empathy, and diagnostic reasoning within a VR environment. A total of 18 volunteers consented to participate in this IRB-approved study at The University of Texas at Tyler, with individual sessions conducted for 15 participants. Thirteen participants completed the post-session survey. 

\subsection{Study Questionnaire}
Participants engaged in a structured questionnaire, adapted from extant simulation assessment literature, to evaluate their prior experience with related technologies, attitudes toward the integration of AI and VR in education, and perceptions of the system's potential as a clinical training tool. The survey consisted of 21 items, including multiple-choice, Likert-scale (rated from 1: Strongly Disagree to 5: Strongly Agree), and open-ended questions. It assessed prior exposure to VR, LLMs, and AI-powered educational tools, while also gauging perceived usefulness and satisfaction with the CLiVR platform. Specific Likert-scale questions required respondents to rate their confidence in using technology for learning, comfort with AI-driven tools, likelihood of adopting CLiVR in teaching physician–patient interactions, and overall satisfaction with the system’s analysis of user communication.

\subsection{Study Results}

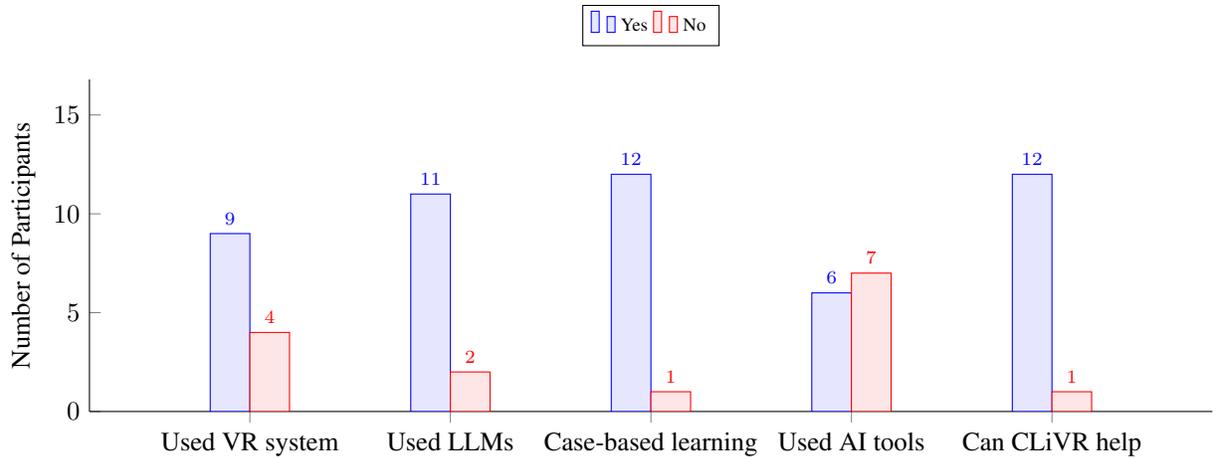
\begin{figure}[!htbp]
\centering
\begin{tikzpicture}
\begin{axis}[
    ybar=0pt,
    bar width=15pt,
    width=\columnwidth,
    height=6cm,
    enlarge x limits=0.2,
    legend style={at={(0.5,-0.15)}, anchor=north, legend columns=-1},
    ylabel={Number of Participants},
    symbolic x coords={
        Used VR system,
        Used LLMs,
        Case-based learning,
        Used AI tools,
        Can CLiVR help
    },
    xtick=data,
    nodes near coords,
    nodes near coords align={vertical},
    ymin=0, ymax=14,
    every node near coord/.append style={font=\scriptsize},
    axis y line*=left,
    axis x line*=bottom,
    enlarge y limits={upper,value=0.2},
    legend style={
    at={(0.5,1.1)},
    anchor=south,
    legend columns=-1,
    font=\scriptsize
},
]

\addplot+[style={fill=blue!10}] 
coordinates {
    (Used VR system,9)
    (Used LLMs,11)
    (Case-based learning,12)
    (Used AI tools,6)
    (Can CLiVR help,12)
};

\addplot+[style={fill=red!10}] 
coordinates {
    (Used VR system,4)
    (Used LLMs,2)
    (Case-based learning,1)
    (Used AI tools,7)
    (Can CLiVR help,1)
};

\legend{Yes, No}
\end{axis}
\end{tikzpicture}
\caption{Yes/No Response Summary (n = 13)}
\end{figure}

\begin{table}[!htbp]
\centering
\begin{tabularx}{\textwidth}{|X|c|c|c|c|}
\hline
\textbf{Survey Item} & \textbf{Count} & \textbf{Mean} & \textbf{Median} & \textbf{Std. Dev.} \\
\hline
Confidence using technology for learning & 13 & 4.08 & 4.0 & 0.64 \\
Comfort with AI-driven learning tools & 13 & 3.77 & 4.0 & 1.17 \\
Likelihood to use for teaching interactions & 13 & 4.00 & 4.0 & 1.22 \\
Likelihood to replace simulated patients & 13 & 2.38 & 2.0 & 1.45 \\
Likelihood to help struggling residents & 13 & 3.54 & 3.0 & 1.27 \\
Satisfaction with analysis detail & 13 & 3.08 & 3.0 & 1.04 \\
\hline
\end{tabularx}
\caption{Descriptive Statistics for Likert-Scale Items (1 = Strongly Disagree, 5 = Strongly Agree)}
\end{table}

A total of 13 participants completed the study survey. Most respondents reported prior experience with virtual reality (9/13) and large language models such as ChatGPT (11/13). Case-based learning was widely used (12/13); however, adoption of AI-powered educational tools—including AR/VR platforms and adaptive simulations—was mixed, with 6 participants reporting prior use and 7 indicating none. Notably, 92.3\% of respondents (12/13) agreed that integrating LLMs with VR would be beneficial for simulating patient–doctor interactions in medical education.

The results derived from the Likert-scale part of our survey for quantitative feedback offer additional insights into user perceptions. Participants expressed a high level of confidence in utilizing technology for educational purposes, with a mean score of 4.08 out of 5, and demonstrated moderate comfort with AI-driven tools, scoring 3.77 out of 5. They were amenable to employing CLiVR for teaching communication skills, with a score of 4.00 out of 5, and for using CLiVR in assisting residents experiencing difficulties with bedside interactions, scoring 3.54 out of 5. However, there was greater reluctance regarding the use of the platform as a complete substitute for simulated patient encounters, as indicated by a score of 2.38 out of 5. Satisfaction with the system’s sentiment analysis and feedback mechanisms averaged 3.08 out of 5. Open-ended responses underscored both strengths and areas for enhancement. Users valued the natural progression of AI-driven dialogue and the realism of the patient avatars, yet identified areas for improvement, such as reducing response latency, incorporating more natural speech patterns, and integrating additional clinical features such as vital signs, laboratory data, and variable patient presentations. Several users cautioned that while CLiVR serves as a valuable adjunct to traditional simulations, it should not be perceived as a complete replacement due to its inherent limitations in capturing the full human nuance of clinical interactions.

We conducted a Wilcoxon signed-rank test to assess whether survey respondents rated their likelihood of using the VR+LLM technology higher than the neutral midpoint of 3 on a 5-point Likert scale. For teaching medical student physician-patient interactions, the median response was significantly greater than the midpoint ($W = 72.5, p = 0.034, r = 0.52$), indicating strong enthusiasm and a large effect size according to Cohen's guidelines. However, participants did not rate the technology as significantly more likely than the midpoint to replace simulated patient encounters ($W = 9.0, p = 0.992, r = -0.71$) or for training residents struggling with bedside interactions ($W = 33.5, p = 0.088, r = -0.23$). These results suggest that while respondents see clear value in using CLiVR in medical education, they remain cautious about fully substituting traditional simulation methods or addressing resident training challenges.

The combined results demonstrate strong enthusiasm for CLiVR as an immersive and scalable communication training platform, with clear recognition of its benefits and current limitations. 

\subsection{Qualitative Feedback Summary} 
The participants identified strengths and areas for improvement of the CLiVR system, offering constructive feedback that underscores its potential in medical education. Many respondents highlighted the system’s applicability to case-based learning, history taking, and structured clinical communication, noting that it could serve as an engaging and repeatable method for practicing essential diagnostic and interpersonal skills. As one participant observed, \textit{This would be useful for history taking and communicating assessments.} An immersive VR environment, coupled with AI-driven conversation, has been seen as a valuable way to expose learners to scenarios that might be rare or logistically challenging in real-world settings. The system was also commended for its capacity to supplement traditional teaching methods and to enhance learners’ engagement. As one respondent observed, \textit{This is an innovative approach that could make clinical skills training more interactive and engaging.} Others mentioned the opportunity to dynamically adapt scenarios, allowing trainees to encounter a variety of patient presentations and communication challenges in a controlled environment. Participants offered suggestions for refinement framed as ways to further increase realism and educational value. These included integrating vital signs, laboratory data, and physical examination findings, as well as diversifying patient profiles to include \textit{individuals with different communication styles or cognitive impairments.} Several studies have also emphasized the importance of expanding cultural and linguistic diversity in the simulated patient population. Overall, respondents recognized CLiVR as a promising supplemental tool rather than a replacement for human-patient encounters that can enrich medical training. Many have emphasized that its greatest value lies in providing a safe, repeatable, and adaptive training environment that complements, but does not substitute for, the irreplaceable human aspect of medicine. These insights will guide future design improvements, including enhanced avatar expressiveness, clinical realism, and the integration of additional training modalities.

\subsection{Emotions of Users in Study}
It is important to note that the emotions of participants in this study were not collected through questionnaires but were automatically inferred from conversation logs. While only \texttt{gemma3} was deployed for real-time monitoring, as described in Section~\ref{sec:sentiment}, the recorded logs were later processed through all models for comparison.

Consistent with the findings from the sentiment analysis module, classification on the unlabeled user-study corpus showed that the models converged strongly toward neutral predictions. Neutral utterances comprised 82\% of all labels for the best BERT model and exceeded 90\% for \texttt{gpt-4o-mini} and \texttt{gemma3n}, producing low entropy values of 0.382–0.576 bits (Table~\ref{tab:entropy}). Inter-model agreement analysis reinforced this trend: the three BERT checkpoints exhibited almost perfect mutual alignment, whereas their alignment with the LLM group was only moderate, indicating divergent decision boundaries in ambiguous contexts.

\begin{table}[!htbp]
\centering
\begin{tabularx}{\columnwidth}{|X|c|c|c|c|}
\hline
\textbf{Model} & \textbf{Negative}
               & \textbf{Neutral}
               & \textbf{Positive}
               & \textbf{Entropy} \\
\hline
bert-base & 0.042 & 0.815 & 0.143 & 0.834 \\
distilbert-base & 0.040 & 0.822 & 0.138 & 0.812 \\
roberta-base & 0.035 & 0.852 & 0.112 & 0.720 \\
mistral & 0.012 & 0.822 & 0.166 & 0.740 \\
gemma3n & 0.035 & 0.896 & 0.068 & 0.576 \\
gemma3 & 0.030 & 0.920 & 0.049 & 0.477 \\
gpt-4o-mini & 0.002 & 0.930 & 0.068 & 0.382 \\
\hline
\end{tabularx}
\caption{Class‑distribution entropy on the 427‑utterance user‑study corpus}
\label{tab:entropy}
\end{table}

\section{Comparison to Existing Simulations}
\label{sec:Comparison}
Commercial VR-based medical training systems, such as Oxford Medical Simulation, PCS Spark, and Body Interact have advanced the accessibility of immersive case-based clinical learning. These platforms typically provide pre-scripted or branched scenarios with structured feedback and a range of clinical cases. However, their conversational flexibility is limited because interactions are often constrained to predefined responses or fixed scenario logic. Recent academic prototypes have begun to integrate AI into VR healthcare simulations; however, most use scripted chatbots or retrieval-based agents rather than LLMs capable of maintaining contextually coherent, dynamic, and persona-driven conversations. Furthermore, while sentiment analysis has been explored in the post-hoc assessment of learner performance, we found no prior VR medical simulation that delivers real-time, in-session empathy feedback during trainee–patient interactions.
To the best of our knowledge, CLiVR is the first VR-based medical communication simulator to integrate symptom-constrained LLM prompting with real-time empathy feedback through sentiment analysis. By grounding patient dialogue in a curated syndrome–symptom knowledge base, CLiVR minimizes hallucinations while maintaining diverse, culturally responsive patient personas. Real-time latency optimization ensures a natural conversation flow, even on standalone VR headsets, supporting deployment in both resource-rich and resource-limited institutions.

\section{Discussions}
\label{sec:Discussion}

In this section, we discuss the limitations and future considerations of LLM and VR integration in medical education based on the feedback received in our exploratory study.

\subsection{Limitations and Feature Recommendations}
Participants in our exploratory study (n=13) recognized CLiVR as an engaging, scalable training tool but noted several areas for improvement. We acknowledge the limited statistical generalizability of our findings, as the study focused on exploratory feasibility rather than definitive efficacy. Future work will involve larger, multi-institutional trials with control conditions (e.g., VR without AI or AI without VR) to isolate each component’s contribution. A key concern was the potential reduction of human connection, with some suggesting AI simulations could “take the personal aspect out of medicine.” Feature recommendations emphasized improving interactivity, realism, and clinical depth. Users proposed reducing response latency for smoother dialogue and integrating vital signs, laboratory data, and physical exam findings for richer reasoning practice. Enhancing speech naturalness and avatar expressiveness was also prioritized to support empathy and engagement. The predominance of “neutral” sentiment classifications reflects typical clinical dialogue but remains useful for detecting empathy shifts or patient distress in high-emotion scenarios.

\subsection{Ethics, Safety, and Deployment}
To ensure the safe and equitable use of CLiVR in medical education, we implemented multiple safeguards against hallucinations and bias in patient representation. All patient scenarios are grounded in a curated syndrome–symptom knowledge base to constrain dialogue within medically plausible boundaries, and system prompts explicitly instruct the model to avoid providing diagnostic conclusions or treatment recommendations beyond its roleplay scope. To reduce demographic bias, the virtual patient library will be expanded to include diverse racial, gender, age, and linguistic profiles, with input from clinical educators and diversity and equity for larger future studies. Finally, we emphasize that CLiVR is not intended to replace SP encounters but to complement them by offering a safe, repeatable, and cost-effective environment for practicing communication skills. 

\section{Conclusion}

CLiVR is an immersive VR system to simulate realistic doctor–patient interactions for medical education. Developed in Unity and deployed on the Meta Quest 3 platform, CLiVR enables speech-based clinical training by leveraging RAG, real-time lip-syncing, and sentiment analysis. Through structured case generation and dynamic conversational roleplay, it supports a scalable and interactive learning experience that complements traditional SP. Our qualitative user study with medical faculty highlighted the system’s strong potential for enhancing communication training, realism, usability, and educational value of the platform, especially in case-based learning and empathy development. The results also indicated high confidence in using technology and a strong willingness to integrate CLiVR into clinical education workflows. Feedback also revealed limitations in avatar realism, diversity and speech naturalness, pointing to areas for technical refinement. Continuous AI and VR technology development promises a path toward more accessible, consistent, and emotionally responsive medical simulations. Future developments will focus on incorporating more diverse patient scenarios, improving the natural dialogue flow, and enhancing multimodal medical record ingestion from diverse sources.

\bibliographystyle{unsrt}
\bibliography{manuscript}
\end{document}